\documentclass[nofootinbib,twocolumn,superscriptaddress,preprintnumbers,longbibliography]{revtex4-1}

\usepackage{amsmath}
\usepackage{amsfonts}
\usepackage{bbm}
\usepackage{microtype}
\usepackage{hyperref}
\usepackage{cleveref}
\usepackage{color}

\newcommand{\dd}{\mathrm{d}}
\newcommand{\sdg}{\sqrt{g}}
\newcommand{\sdf}{\sqrt{f}}
\newcommand{\mn}{{\mu\nu}}

\newcommand{\eff}{\mathrm{eff}}
\newcommand{\Mp}{M_\mathrm{Pl}}
\newcommand{\mfp}{m_\mathrm{FP}}
\newcommand{\Mf}{M_f}

\newcommand{\para}[1]{\pdfbookmark[1]{#1}{#1}\par\vspace{2mm}\noindent{\bf\emph{{#1}}}.---}

\makeatletter
\DeclareRobustCommand{\rcite}[1]{%
  \rcite@aux#1,\@nil{#1}%
}
\def\rcite@aux#1,#2\@nil#3{%
  \if\relax#2\relax
    Ref.~\cite{#3}%
  \else
    Refs.~\cite{#3}%
  \fi
}
\makeatother

\begin{document}

\title{Bimetric gravity is cosmologically viable}
\author{Yashar Akrami}
\email{akrami@thphys.uni-heidelberg.de}
\affiliation{Nordita, KTH Royal Institute of Technology and Stockholm University\\Roslagstullsbacken 23, SE-10691 Stockholm, Sweden}
\affiliation{Institut f\"ur Theoretische Physik, Ruprecht-Karls-Universit\"at Heidelberg\\Philosophenweg 16, 69120 Heidelberg, Germany}
\author{S.F. Hassan}
\email{fawad@fysik.su.se}
\affiliation{Nordita, KTH Royal Institute of Technology and Stockholm University\\Roslagstullsbacken 23, SE-10691 Stockholm, Sweden}
\affiliation{Department of Physics and the Oskar Klein Centre, Stockholm University\\
AlbaNova University Center, SE 106 91 Stockholm, Sweden}
\author{Frank K\"onnig}
\email{koennig@thphys.uni-heidelberg.de}
\affiliation{Nordita, KTH Royal Institute of Technology and Stockholm University\\Roslagstullsbacken 23, SE-10691 Stockholm, Sweden}
\affiliation{Institut f\"ur Theoretische Physik, Ruprecht-Karls-Universit\"at Heidelberg\\Philosophenweg 16, 69120 Heidelberg, Germany}
\author{Angnis Schmidt-May}
\email{angniss@itp.phys.ethz.ch}
\affiliation{Nordita, KTH Royal Institute of Technology and Stockholm University\\Roslagstullsbacken 23, SE-10691 Stockholm, Sweden}
\affiliation{Institut f\"ur Theoretische Physik, Eidgen\"ossische Technische Hochschule Z\"urich\\Wolfgang-Pauli-Strasse 27, 8093 Z\"urich, Switzerland}
\author{Adam R. Solomon}
\email{a.r.solomon@damtp.cam.ac.uk}
\thanks{Corresponding author}
\affiliation{Nordita, KTH Royal Institute of Technology and Stockholm University\\Roslagstullsbacken 23, SE-10691 Stockholm, Sweden}
\affiliation{Institut f\"ur Theoretische Physik, Ruprecht-Karls-Universit\"at Heidelberg\\Philosophenweg 16, 69120 Heidelberg, Germany}
\affiliation{DAMTP, Centre for Mathematical Sciences, University of Cambridge\\ Wilberforce Rd., Cambridge CB3 0WA, UK}

\begin{abstract} Bimetric theory describes gravitational interactions in the presence of an extra spin-2 field. Previous work has suggested that its cosmological solutions are generically plagued by instabilities. We show that by taking the Planck mass for the second metric, $M_f$, to be small, these instabilities can be pushed back to unobservably early times. In this limit, the theory approaches general relativity with an effective cosmological constant which is, remarkably, determined by the spin-2 interaction scale. This provides a late-time expansion history which is extremely close to $\Lambda$CDM, but with a technically-natural value for the cosmological constant. We find $M_f$ should be no larger than the electroweak scale in order for cosmological perturbations to be stable by big-bang nucleosynthesis. We further show that in this limit the helicity-0 mode is no longer strongly-coupled at low energy scales.
\end{abstract}

\keywords{modified gravity, massive gravity, background cosmology, cosmic acceleration, dark energy, bimetric gravity, bigravity}
\preprint{NORDITA-2015-31}

\maketitle

\begin{quote}``The reports of my death have been greatly exaggerated.''

---\textit{Metrics Twain}
\end{quote}

\para{Introduction} The Standard Model of particle physics contains fields with spins 0, 1/2, and 1, describing matter as well as the strong and electroweak forces. General relativity (GR) extends this to the gravitational interactions by introducing a massless spin-2 field. There is theoretical and observational motivation to seek physics beyond the Standard Model and GR. In particular, GR is nonrenormalizable and is associated with the cosmological constant, dark energy, and dark matter problems. To compound the puzzle, the GR-based $\Lambda$-cold dark matter ($\Lambda$CDM) model provides a very good fit to observational data, despite its theoretical problems. In order to be observationally viable, any modified theory of gravity must be able to mimic GR over a wide range of distances.

A natural possibility for extending the set of known classical field
theories is to include additional spin-2 fields and interactions.
While ``massive'' and ``bimetric'' theories of gravity have a long
history \cite{Fierz:1939ix,Isham:1971gm}, nonlinear
theories of interacting spin-2 fields were found, in general, to suffer from the
Boulware-Deser (BD) ghost instability \cite{Boulware:1973my}. Recently a
particular bimetric theory (or bigravity) has been shown to avoid this
ghost instability~\cite{Hassan:2011zd,Hassan:2011ea}. This theory
describes nonlinear interactions of the gravitational metric with an
additional spin-2 field. It is an extension of an earlier ghost-free theory
of massive gravity (a massive spin-2 field on a nondynamical flat
background) \cite{Creminelli:2005qk,deRham:2010ik,deRham:2010kj} for
which the absence of the BD ghost at the nonlinear level was established in
\rcite{Hassan:2011hr,Hassan:2011tf,Hassan:2011ea,Hassan:2012qv}.

Including spin-2 interactions modifies GR, \textit{inter alia}, at large distances. Bimetric theory is therefore a candidate to explain the accelerated expansion of the Universe \cite{Riess:1998cb,Perlmutter:1998np}. Indeed, bigravity has been shown to possess Friedmann-Lema\^{\i}tre-Robertson-Walker (FLRW) solutions which can match observations of the cosmic expansion history, even in the absence of vacuum energy \cite{Volkov:2011an,Comelli:2011zm,vonStrauss:2011mq,Akrami:2012vf,Akrami:2013pna,Konnig:2013gxa,Enander:2014xga}.\footnote{Stable FLRW solutions do not exist in massive gravity \cite{D'Amico:2011jj,Gumrukcuoglu:2011ew,Gumrukcuoglu:2011zh,Vakili:2012tm,DeFelice:2012mx,Fasiello:2012rw,DeFelice:2013awa}.} Linear perturbations around these cosmological backgrounds have also been studied extensively \cite{Berg:2012kn,Comelli:2012db,Konnig:2014dna,Solomon:2014dua,Konnig:2014xva,Lagos:2014lca,Cusin:2014psa,Yamashita:2014cra,DeFelice:2014nja,Fasiello:2013woa,Enander:2015vja,Amendola:2015tua,Johnson:2015tfa,Konnig:2015lfa}. The epoch of acceleration is set by the mass scale $m$ of the spin-2 interactions. Unlike a small vacuum energy, $m$ is protected from large quantum corrections due to an extra diffeomorphism symmetry that is recovered in the limit $m\to0$, just as fermion masses are protected by chiral symmetry in the Standard Model (see \rcite{deRham:2013qqa} for an explicit analysis in the massive gravity setup). This makes interacting spin-2 fields especially attractive from a theoretical point of view.

Cosmological solutions lie on one of two branches, called the finite
and infinite branches.\footnote{There is a third branch containing
  bouncing solutions, but these tend to have pathologies
  \cite{Konnig:2015lfa}.} The infinite-branch models can have
sensible backgrounds \cite{Konnig:2013gxa,Konnig:2014xva}, but the
perturbations have been found to contain ghosts in both the scalar and
tensor sectors \cite{Lagos:2014lca,Cusin:2014psa,Konnig:2015lfa}.
Most viable background solutions lie on the finite branch
\cite{vonStrauss:2011mq,Akrami:2012vf,Akrami:2013pna,Konnig:2013gxa}.
While these avoid the aforementioned ghosts, they contain a scalar
instability at early times
\cite{Comelli:2012db,Konnig:2014xva,Lagos:2014lca} that invalidates
the use of linear perturbation theory and could potentially rule these
models out. For parameter values thought to be favored by data, this
instability was found to be present until recent times
(i.e., a similar time to the onset of cosmic acceleration) and thus
seemed to spoil the predictivity of bimetric cosmology.

In this Letter we study a physically well-motivated region in the
parameter space of bimetric theory that has been missed in earlier
work due to a ubiquitous choice of parameter rescaling. We demonstrate how in
this region the instability problem in the finite branch can be
resolved while the model still provides late-time acceleration in
agreement with observations.

Our search for viable bimetric cosmologies will be guided by the
precise agreement of GR with data on all scales, which motivates us to
study models of modified gravity which are close to their GR limit.
Often this limit is dismayingly trivial; if a theory of modified
gravity is meant to produce late-time self-acceleration in the absence
of a cosmological constant degenerate with vacuum energy, then we
would expect that self-acceleration to disappear as the theory
approaches GR. We will see, however, that there exists a GR limit of bigravity
which retains its self-acceleration, leading to a GR-like universe
with an effective cosmological constant produced purely by the spin-2
interactions.

\para{Bimetric gravity} The ghost-free action for bigravity
containing metrics $g_\mn$ and $f_\mn$ is given by
\cite{Hassan:2011zd,Hassan:2011vm}
\begin{align}
   S = \int \dd^4x\Big[&-\frac{\Mp^2}{2}\sdg R(g) -\frac{\Mf^2}{2}
   \sdf R(f) \nonumber \\ 
   & + m^2\Mp^2\sdg\, V(\mathbb X)+
   \sdg\mathcal{L}_m\left(g, \Phi_i\right)\Big].
\label{eq:action} 
\end{align}
Here $\Mp$ and $\Mf$ are the Planck masses for $g_\mn$ and $f_\mn$,
respectively, and we will frequently refer to their ratio, 
\begin{equation}
\alpha \equiv \frac \Mf\Mp .
\end{equation}
The potential $V(\mathbb X)$ is constructed from the elementary
symmetric polynomials $e_n(\mathbb X)$ of the eigenvalues of the
matrix $\mathbb X \equiv \sqrt{g^{-1}f}$, defined by 
\begin{equation}
\mathbb X^\mu{}_\alpha \mathbb X^\alpha{}_\nu \equiv
g^{\mu\alpha}f_{\alpha\nu}, 
\end{equation}
and has the form
\cite{deRham:2010kj,Hassan:2011vm},\footnote{This is a
    generalization of the massive-gravity potential
    \cite{deRham:2010kj} (to which it reduces for
    $f_{\mu\nu}=\eta_{\mu\nu}$ and a restricted set of $\beta_n$)
    given in \rcite{Hassan:2011vm}.}  
\begin{equation}
\sdg\, V(\mathbb X) = \sdg\beta_0 + \sdg
\sum_{n=1}^{3}\beta_ne_n(\mathbb X) + \sdf\beta_4. 
\end{equation}
In the above, $m$ is a mass scale and $\beta_n$ are dimensionless
interaction parameters. $\beta_0$ and $\beta_4$ parameterize the
vacuum energies in the two sectors. Guided by the absence of ghosts
and the weak equivalence principle, we take the matter sector to be
coupled to $g_\mn$.\footnote{More general matter couplings not
  constrained by these requirements have been studied in
  \rcite{Akrami:2013ffa,Tamanini:2013xia,Akrami:2014lja,Yamashita:2014fga,deRham:2014naa,deRham:2014fha,Noller:2014sta,Enander:2014xga,Solomon:2014iwa,Schmidt-May:2014xla,Heisenberg:2014rka,Gumrukcuoglu:2014xba,Noller:2014ioa,Gumrukcuoglu:2015nua,Comelli:2015pua,Hinterbichler:2015yaa}.}
Then the vacuum-energy contributions from the matter sector
$\mathcal{L}_\mathrm{m}$ are captured in $\beta_0$. We can interpret
$g_\mn$ as the spacetime metric used for measuring distance and time,
while $f_\mn$ is an additional symmetric tensor that mixes
nontrivially with gravity. As we discuss further below, the two
metrics do not correspond to the spin-2 mass eigenstates but each
contain both massive and massless components. Even before fitting to
observational data, the parameters in the bimetric action are subject
to several theoretical constraints. For instance, the squared mass of
the massive spin-2 field needs to be positive, it must not violate the
Higuchi bound \cite{Higuchi:1986py,Higuchi:1989gz}, and ghost modes
should be absent.

In terms of the Einstein tensor, $G_\mn$, the equations of motion for 
the two metrics take the form  
\begin{align}
G_\mn(g) + m^2V^g_\mn &= \frac{1}{\Mp^2}T_\mn, \label{eq:einsteing} \\
\alpha^2 G_\mn(f) + m^2V^f_\mn &= 0, \label{eq:einsteinf}
\end{align}
where $V^{(g,f)}_\mn$ are determined by varying the interaction
potential, $V$. Taking the divergence of \cref{eq:einsteing}
and using the Bianchi identity leads to the \emph{Bianchi constraint},  
\begin{equation}
\nabla_{(g)}^\mu V^g_\mn = 0. \label{eq:bianchi}
\end{equation}
The analogous equation for $f_\mn$ carries no additional
information due to the general covariance of the action.

Finally, note that the action \eqref{eq:action} has a status similar
to Proca theory on curved backgrounds. It is therefore expected to require
an analogue of the Higgs mechanism, with new degrees of freedom, in order to have
improved quantum behavior. The search for a ghost-free Higgs mechanism for gravity is still in progress \cite{Goon:2014paa}.

\para{The GR limit} When bigravity is linearized around proportional
backgrounds $\bar f_\mn=c^2\bar g_\mn$ with constant
$c$,\footnote{These correspond to Einstein spaces and, for
  nonvanishing $\alpha$, solve the field equations only in vacuum. A
  quartic equation determines $c=c(\beta_n,\alpha)$.}   
\begin{align}
g_\mn &= \bar g_\mn + \frac{1}{\Mp}\delta g_\mn, \\
f_\mn &= c^2\bar g_\mn + \frac{c}{\Mf}\delta f_\mn,
\end{align}
the canonically-normalized perturbations can be diagonalized into
massless modes $\delta G_\mn$ and massive modes $\delta M_\mn$ as
\cite{Hassan:2011zd,Hassan:2012wr} 
 \begin{align}
\delta G_\mn &\propto \left(\delta g_\mn+c\alpha\,\delta f_\mn\right), \\
\delta M_\mn &\propto \left(\delta f_\mn-c\alpha\,\delta g_\mn\right).
\end{align}
Notice that when $\alpha\to0$ (or $\Mp\gg\Mf$), the massless state 
aligns with $\delta g_\mn$, i.e., up to normalization,  
\begin{equation}
\delta G_\mn\to \delta g_\mn + \mathcal{O}(\alpha^2).
\end{equation}
Because $g_\mn$ is the physical metric, this suggests that
$\alpha\to0$ is the general-relativity limit of bigravity.\footnote{See \rcite{Berezhiani:2007zf} for an early discussion of such a limit.} We will see
below that the nonlinear field equations indeed reduce to Einstein's
equations for $\alpha=0$ and that the limit is continuous. Thus
$g_\mn$ is close to a GR solution for sufficiently small values of $\alpha$.
We therefore identify $\Mp$ with the measured physical Planck mass
whenever $\alpha\ll 1$, holding it fixed while making $M_f$ smaller.
Interestingly, in the bimetric setup a large physical Planck mass is
correlated with the fact that gravity is approximated well by a
massless field. In other words, when bimetric theory is close to GR,
the gravitational force is naturally weak.

The GR limit can be directly realized at the nonlinear level
\cite{Baccetti:2012bk,Hassan:2014vja}. The metric potentials satisfy 
the identity  
\begin{equation}
\sdg g^{\mu\alpha}V^g_{\alpha\nu} + \sdf f^{\mu\alpha}V^f_{\alpha\nu}
= \sdg V\delta^\mu{}_\nu, \label{eq:identity} 
\end{equation}
where $V$ is the potential in the action \eqref{eq:action}. 
For $M_f=0$, the $f_\mn$ equation \eqref{eq:einsteinf} gives 
$V^f_\mn=0$, an algebraic constraint on $f_\mn$. Then, using the above identity, the $g_\mn$ equation
\eqref{eq:einsteing} becomes
\begin{equation}
G_\mn(g) + m^2Vg_\mn = \frac{1}{\Mp^2}T_\mn.
\label{eq:combeq} 
\end{equation}
Since $T_\mn$ is conserved, taking the divergence gives
\begin{equation}
\partial_\mu V = 0.
\label{eq:delV}
\end{equation}
We see that \cref{eq:combeq} is the Einstein equation for $g_\mn$ with
cosmological constant $m^2V$. Remarkably, because $V$ depends on $f_\mn$
and all the $\beta_n$, this effective cosmological constant is
generically \emph{not} simply the vacuum energy from matter loops
(which is parameterized by $\beta_0$). Even in the GR limit, the
impact of the spin-2 interactions remains and bigravity's
self-acceleration survives.  

It is straightforward to see that, unlike the $m\rightarrow 0$ limit,
the $\alpha\rightarrow 0$ limit is not affected by the van
Dam-Veltman-Zakharov (vDVZ) discontinuity
\cite{vanDam:1970vg,Zakharov:1970cc}. The cause of this discontinuity
is the Bianchi constraint \eqref{eq:bianchi} which constrains the
solutions even when $m=0$. On the contrary, when $\alpha\rightarrow
0$, the Bianchi constraint simply reduces to \cref{eq:delV} and is
automatically satisfied.

The conditions $V^f_\mn=0$ and $\partial_\mu V=0$ determine $f_\mn$
algebraically in terms of $g_\mn$, generically as $f_\mn=c^2g_\mn$. In
the limit $\Mf=0$, the $f$ sector is infinitely strongly
coupled.\footnote{Strongly-coupled gravity in the context of GR has
  been studied, for instance, in
  \rcite{Henneaux:1981su,Francisco:1984jp,Salopek:1998un,Niedermaier:2014ywa}
  and has been argued to allow for a simplified quantum-mechanical
  treatment.} Due to the nontrivial potential, this causes the $f$
metric to exactly follow the $g$ metric (both at the background and
perturbative levels), while the $g$ sector remains weakly coupled.

\para{Strong-coupling scales} We now argue that at energy scales relevant to cosmology, this model avoids known strong-coupling issues, sometimes contrary to intuition gained from massive gravity.

There are several strong-coupling scales one might expect to arise. At an energy scale $k$, the $f$ sector has an effective coupling $k/\Mf$,
as can be seen from expanding the Einstein-Hilbert action in $\delta
f_\mn/\Mf$, just as in GR. Then, for small but nonzero $\alpha$, which
is the case of interest here, one might worry that perturbations of
$f_\mn$ with momentum $k$ become strongly coupled at low scales $k\sim
M_f$.  However, we have seen that in the limit of infinite strong
coupling, $\Mf=0$, $f_\mn$ becomes nondynamical and is entirely
determined in terms of $g_\mn$, while the $g_\mn$ equation is
degenerate with GR and its perturbations remain weakly coupled. Due to
the continuity of the limit, we expect that, for small enough
$\alpha$, strong-coupling effects will continue to not affect the $g$ sector, even
when perturbations of $f_\mn$ are strongly coupled at relatively small
energy scales. In practice, however, since the measured value of $\Mp$
is very large, even reasonably high values of $\Mf$ can still lead to
small $\alpha$.  In cosmological applications, all observable
perturbations satisfy $k/M_f\ll 1$ for $M_f\gg100H_0\sim10^{-31}$~eV, roughly the
scale at which linear cosmological perturbation theory breaks down at
recent times, so that perturbations of $f_\mn$ remain weakly coupled
in any case.

Another potentially-problematic scale is associated with the helicity-0 mode of
the massive graviton. In massive gravity, this mode becomes strongly coupled at
the scale \cite{ArkaniHamed:2002sp,Schwartz:2003vj}
\begin{equation}
\Lambda_3 \equiv \left(m^2\Mp\right)^{1/3},
\label{eq:mgrL3}
\end{equation}
where $m$ is defined to coincide with the Fierz-Pauli mass \cite{Fierz:1939ix} on flat backgrounds. This scale is rather small,
$\Lambda_3\sim10^{-13}~\mathrm{eV}\sim(1000~\mathrm{km})^{-1}$ for
$m\sim H_0\sim10^{-33}$~eV, and severely restricts the applicability of
massive gravity \cite{Burrage:2012ja}. The same scale also appears in
the decoupling-limit analysis of bimetric theory
\cite{Fasiello:2013woa}, where $m$ is now the parameter in front of the
potential in the action \eqref{eq:action}. In the limit $\alpha\to0$,
the $f$ sector approaches massive gravity \cite{Hassan:2014vja} and
one might worry that the strong-coupling problem persists or becomes
worse with the emergence of an even lower scale $(m^2\Mf)^{1/3}$.  This is not the case. In the bimetric context, the scale defined in \cref{eq:mgrL3} is not physical, since $m^2$ is
degenerate with the $\beta_n$. The physically relevant strong-coupling
scale must be defined with respect to the bimetric Fierz-Pauli mass
\cite{Hassan:2012wr},
\begin{equation}
\mfp^2=m^2\left(\frac{1}{c^2\alpha^2}+1\right)
\left(c\beta_1+2c^2\beta_2+c^3\beta_3\right),
\label{eq:mfp}
\end{equation}
which is only defined around proportional backgrounds, $f_\mn = c^2 g_\mn$. In the massive-gravity limit, $\alpha\rightarrow\infty$, the helicity-0
mode is mostly contained in $g$ with a strong-coupling scale
\begin{equation}
\Lambda_3 \equiv \left(\mfp^2\Mp\right)^{1/3},
\label{eq:bimL3}
\end{equation}
consistent with \cref{eq:mgrL3} for appropriately restricted
parameters. However, in the GR limit, $\alpha\rightarrow 0$, the
helicity-0 mode resides mostly in $f$, where the strong-coupling scale is
\begin{equation}
\tilde\Lambda_3 \equiv \left(\mfp^2 \Mf\right)^{1/3} \rightarrow
\left(\frac{m^2 \Mp}{\alpha}\, {\mathcal O}(\beta_n)\right)^{1/3}, 
\end{equation}
which is no longer small. Note that for solutions that admit this limit, $c$ 
becomes independent of $\alpha$. We can also consider the $\alpha\rightarrow 0$ limit of
\cref{eq:bimL3}, to verify that the small part of the helicity-0 mode
in $g$ is not strongly coupled,
\begin{equation}
\Lambda_3 \rightarrow 
\left(\frac{m^2 \Mp}{\alpha^2}\,{\mathcal O}(\beta_n)\right)^{1/3}.
\end{equation} 
This is even higher than $\tilde\Lambda_3$. Therefore the strong-coupling issues with the helicity-0 mode are alleviated, rather than exacerbated, when $\alpha\to0$. 

\para{Cosmology} We now proceed to apply the above arguments to the
particular example of a homogeneous and isotropic universe. We will
take both metrics to be of the diagonal FLRW form \cite{Volkov:2011an,
vonStrauss:2011mq, Comelli:2011zm},\footnote{See \rcite{Nersisyan:2015oha} and the references therein for
  other possible metrics in bimetric cosmology.}
\begin{align}
g_\mn \dd x^\mu \dd x^\nu &= -\dd t^2 + a^2(t)\delta_{ij}\dd x^i \dd x^j, \\
f_\mn \dd x^\mu \dd x^\nu &= -X^2(t)\dd t^2 + Y^2(t)\delta_{ij}\dd x^i \dd x^j,
\end{align}
where we can freely choose the cosmic-time coordinate for $g_\mn$ ($g_{00}=-1$) because of general covariance. Because matter couples minimally to $g_\mn$, this
choice is physical, and $a(t)$ corresponds to the scale factor
inferred from observations. We furthermore take the matter source to
be a perfect fluid, $T^{\mu}_{~\nu}=\mathrm{diag}(-\rho,p,p,p)$. The
$g$-metric equation \eqref{eq:einsteing} leads to the Friedmann
equation,  
\begin{equation}
3H^2 = \frac{\rho}{\Mp^2} + m^2\left(\beta_0 + 3\beta_1y + 3\beta_2y^2
+ \beta_3y^3\right), \label{eq:fried} 
\end{equation}
where the Hubble rate is defined as $H\equiv \dot a/a$ and the ratio
of the scale factors is 
\begin{equation}
y\equiv \frac Ya.
\end{equation}
The analogous equation for the $f$ metric is
\begin{equation}
3K^2 = \frac{m^2}{\alpha^2}X^2\left(\frac{\beta_1}{y^3} +
3\frac{\beta_2}{y^2} + 3\frac{\beta_3}{y}+\beta_4\right),
\label{eq:friedf}  
\end{equation}
with $K\equiv \dot Y/Y$. The final ingredient is the Bianchi
constraint \eqref{eq:bianchi}, which yields 
\begin{equation}
\left(HX-Ky\right)\left(\beta_1+2\beta_2y+\beta_3y^2\right) =0. 
\label{eq:bianchifrw} 
\end{equation}
Taking the first or second term of \cref{eq:bianchifrw} to vanish
selects the so-called dynamical or algebraic branches, respectively.
Perturbations in the algebraic branch are pathological
\cite{Comelli:2012db}, so we will consider the dynamical branch in
which the $f$-metric lapse is fixed,   
\begin{equation}
X = \frac{Ky}{H}.
\end{equation}
Inserting this into the $f_\mn$ equation \eqref{eq:friedf}
transforms it into an ``alternate'' Friedmann equation, 
\begin{equation}
3\alpha^2 H^2 = m^2\left(\frac{\beta_1}{y} + 3\beta_2 + 3\beta_3y
+\beta_4y^2\right). \label{eq:friedalt} 
\end{equation}
We take at least two of the $\beta_n$ for $n\geq1$ to be nonzero in
order to ensure the existence of interesting solutions in the GR limit
$\alpha\to0$. The solutions to \cref{eq:friedalt} in the GR limit are
always on the ``finite'' branch, i.e., $y$ evolves from 0 to a
finite late-time value. The perturbations on this branch are healthy
\emph{except} for a scalar instability, which we discuss below. 

\Cref{eq:friedalt} has two features which
are useful for our purposes. First, in the limit $\alpha\to0$ it tends
to a polynomial constraint that leads to a constant solution for $y$,
so that the potential term in the Friedmann equation \eqref{eq:fried}
becomes a cosmological constant. This provides an explicit example of
the statement above that as $\alpha\to0$, the theory approaches
general relativity with an effective cosmological constant (even with
$\beta_0=0$). Recall that even though the theory approaches GR in this
limit, the bigravity interactions survive in the form of this
constant. The other useful feature is that, because \cref{eq:friedalt} 
does not involve $\rho$, it can be used to rephrase the potential term
in \cref{eq:fried} in terms of the Hubble rate. This will allow us to
determine the time-dependence of the potential term order by order in
$\alpha$.\footnote{One can also combine \cref{eq:fried,eq:friedalt} to
  obtain a quartic equation for $y$ involving $\rho$
  \cite{Volkov:2011an,vonStrauss:2011mq,Comelli:2011zm,Akrami:2012vf,Solomon:2014dua},
  but this is more cumbersome as it involves higher powers of $y$ than
  \cref{eq:friedalt} does.}

\para{The effective cosmological constant} Let us illustrate the new
viable bimetric cosmologies qualitatively by selecting the model with
$\beta_0 = \beta_3 = \beta_4 = 0$,\footnote{Since we are interested in
  finding self-accelerating solutions in the absence of vacuum energy,
  we will set $\beta_0=0$ herein, but emphasize that this is not
  necessary.} which we will refer to as the $\beta_1\beta_2$ model.
The Friedmann and ``alternate'' Friedmann 
equations \eqref{eq:fried} and \eqref{eq:friedalt} are 
\begin{align}
3H^2 &= \frac{\rho}{\Mp^2} + 3m^2\left(\beta_1y +
\beta_2y^2\right), \label{eq:b1b2fried} \\ 
3\alpha^2H^2 &= m^2 \left(\frac{\beta_1}{y} +
3\beta_2\right). \label{eq:b1b2friedalt} 
\end{align}
We can use \cref{eq:b1b2friedalt} to eliminate $y$ in \cref{eq:b1b2fried}. It is instructive to work in the GR limit where
\cref{eq:b1b2friedalt} gives 
\begin{equation}
y \xrightarrow{\alpha\to0}-\frac13\frac{\beta_1}{\beta_2}.
\end{equation}
The $\alpha\to0$ limit is nonsingular only if both $\beta_1$ and
$\beta_2$ are nonzero. Plugging this into \cref{eq:b1b2fried} we obtain   
\begin{equation}
3H^2 = \frac{\rho}{\Mp^2} - \frac23\frac{\beta_1^2}{\beta_2}
m^2. \label{eq:b1b2grfried} 
\end{equation}
The effective cosmological constant is
\begin{equation}
\Lambda_\eff = -\frac23\frac{\beta_1^2}{\beta_2}m^2.
\end{equation}
Late-time acceleration requires $\beta_2<0$.

When we are not exactly in the GR limit, we should consider
corrections to \cref{eq:b1b2grfried}, 
\begin{align}
3H^2 &= \frac{\rho}{\Mp^2} +\frac{\beta_1^2m^4}{3\left(H^2\alpha^2-
\beta_2m^2\right)^2}\left(3\alpha^2H^2-2\beta_2m^2\right)
\nonumber \\ 
&= \frac{\rho}{\Mp^2} - \frac23\frac{\beta_1^2}{\beta_2} m^2 -
\frac{\alpha^2\beta_1^2}{3\beta_2^2}H^2 + \mathcal{O}(\alpha^4). 
\end{align}
This expansion is valid as long as
\begin{equation}
 H^2 \lesssim \frac{\beta_2m^2}{\alpha^2}. \label{eq:backvalid}
\end{equation}
Rearranging and again keeping terms up to $\mathcal{O}(\alpha^2)$, we
find a standard Friedmann equation with a time-varying effective
cosmological constant given by 
\begin{equation}
\Lambda_\eff = -\frac23\frac{\beta_1^2}{\beta_2}m^2 -\frac29\frac{\beta_1^2}{\beta_2^2}\alpha^2\left(\frac{\rho}{2\Mp^2} - \frac{\beta_1^2}{3\beta_2}m^2\right) + \mathcal{O}(\alpha^4).
\end{equation}
Because matter is coupled minimally to $g_\mn$, it will have the
standard behavior $\rho \sim a^{-3(1+w)}$, where $w=p/\rho$ is the
equation-of-state parameter, allowing $\rho$ to stand in for time.
This captures the first hint of the dynamical dark energy that is
typical of bigravity
\cite{vonStrauss:2011mq,Akrami:2012vf,Akrami:2013pna,Konnig:2013gxa,Enander:2014xga}. 

\begin{table}
\begin{ruledtabular}
\begin{tabular}{c|c|c}
Model & $\Lambda_\eff$ ($\alpha\to0$) & $\mathcal{O}(\alpha^2)$ correction \\ \hline
$\beta_1,\beta_2\neq0$ & $-\frac23\frac{\beta_1^2}{\beta_2}m^2$ & $-\frac29\frac{\beta_1^2}{\beta_2^2}\alpha^2\left(\frac{\rho}{2\Mp^2} - \frac{\beta_1^2}{3\beta_2}m^2\right)$ \\
$\beta_1,\beta_3\neq0$ & $\frac{8}{3\sqrt3}\frac{\beta_1^{3/2}}{\sqrt{-\beta_3}}m^2$ & $\frac{\beta_1}{\beta_3}\alpha^2\left(\frac{\rho}{3\Mp^2} - \frac{8\beta_1^{3/2}}{9\sqrt{-3\beta_3}}m^2 \right)$ \\
$\beta_1,\beta_4\neq0$ & $3 \frac{\beta _1^{4/3}}{\sqrt[3]{-\beta _4}} m^2$ & $-\left(-\frac{\beta_1}{\beta_4}\right)^\frac23\alpha^2\left(\frac{\rho}{\Mp^2} + 3\frac{\beta_1^{4/3}}{\sqrt[3]{-\beta_4}}m^2\right)$ \\
$\beta_2,\beta_3\neq0$ & $2\frac{\beta _2^3}{\beta _3^2} m^2$ & $-\frac{\beta_2^2}{\beta_3^2}\alpha^2\left(\frac{\rho}{\Mp^2}+\frac{2 \beta _2^3}{\beta _3^2} m^2\right)$ \\
$\beta_2,\beta_4\neq0$ & $-9\frac{\beta_2^2}{\beta_4}m^2$ & $3\frac{\beta_2}{\beta_4}\alpha^2\left(\frac{\rho}{\Mp^2}-\frac{9\beta_2^2}{\beta_4}m^2\right)$
\end{tabular}
\caption{The effective cosmological constant and lowest-order
  corrections (which are time-dependent through $\rho$) for a variety
  of two-parameter models. We have chosen solution branches which lead
  to positive $\Lambda_\eff$ for appropriate signs of the $\beta_n$,
  and generally take $\beta_1\geq0$ based on viability conditions
  \cite{Konnig:2013gxa}. The $\beta_3,\beta_4\neq0$ model does not
  possess a finite-branch solution \cite{Konnig:2013gxa}.} 
\label{tab:eff-cc}
\end{ruledtabular}
\end{table}

These results generalize easily to other parameter combinations. We
list the effective cosmological constant up to $\mathcal{O}(\alpha^2)$
for all the two-parameter models (setting $\beta_0=0$) in
\cref{tab:eff-cc}. We remind the reader that, in order for the
$\alpha\to0$ limit to be well-behaved, at least two of the $\beta_n$
parameters (excluding the vacuum energy contribution, $\beta_0$) must
be nonzero. 

\para{Exorcising the instability} The stability of cosmological
perturbations in bigravity was investigated in \rcite{Konnig:2014xva}
by determining the full solutions to the linearized Einstein equations
in the subhorizon r\'egime. The perturbations were shown to obey a WKB
solution given by 
\begin{equation}
\Phi \sim e^{i\omega N},
\end{equation}
where $\Phi$ represents any of the scalar perturbation
variables, $N\equiv\ln a$, and we have taken the limit $k\gg aH$ where
$k$ is the comoving wavenumber. The eigenfrequencies $\omega$ were
presented for particular models in \rcite{Konnig:2014xva}, where it
was found that all models with viable backgrounds have $\omega^2<0$ at
early times, revealing a gradient instability that only ends at a very
low redshift. Using the formulation of the linearized equations of
motion presented in \rcite{Lagos:2014lca}, we can write the
eigenfrequencies for general $\beta_n$ and $\alpha$ in the compact
form \cite{Konnig:2015lfa}
\begin{align}
\left(\frac{aH}{k}\right)^2\omega^2 &= 1 + \frac{\left(\beta _1+4 \beta _2 y+3 \beta _3 y^2\right) y'}{3y
   \left(\beta _1+2 \beta _2 y+\beta _3 y^2\right)} \nonumber\\
   &\hphantom{{}=}- \frac{\left(1+\alpha ^2 y^2\right) \left(\beta_1-\beta_3 y^2\right) y'^2}{3\alpha ^2 y^3\tilde\rho  (1+w) }, \label{eq:eigenfreqs}
\end{align}
where $\tilde\rho\equiv \rho/m^2\Mp^2$ and primes denote $\dd/\dd\ln a$.

We apply this to the $\beta_1\beta_2$ model. Assuming a
  universe dominated by dust ($w=0$), $\omega^2$ crosses zero when\footnote{We have used \cref{eq:b1b2fried,eq:b1b2friedalt} and their derivatives to solve for $y'$ and $\rho$ in \cref{eq:eigenfreqs} in terms of $\beta_n$ and $y$ \cite{Solomon:2014dua}. Note that $\omega^2=0$ does not imply strong coupling because, while the gradient terms vanish, the kinetic terms remain nonzero.}
\begin{align}
18 \alpha ^2\beta_2 \left(\alpha ^2 \beta _1^2 +4  \beta _2^2\right)y^5+9 \alpha^2\beta_1\left(\alpha^2\beta _1^2+10  \beta_2^2\right)y^4 \nonumber\\
   +48 \alpha ^2 \beta _1^2 \beta _2 y^3+6 \beta_2\left(2 \alpha ^2 \beta _1^2- \beta _2^2\right) y^2-6 \beta _1^2 \beta _2 y -\beta _1^3=0.
\end{align}
Solving this for $y$, we can then use \cref{eq:b1b2friedalt} to
determine the value of Hubble rate at the \emph{transition era},
before which the gradient instability is present and after which it
vanishes. While this solution is too complicated to write down
explicitly, in the limit $\alpha\to0$ the leading-order term is
remarkably simple,\footnote{While \cref{eq:b1b2hstab} only holds exactly in the presence of dust, $w=0$, for other reasonable equations of state, such as radiation ($w=1/3$), it will only be modified by an $\mathcal{O}(1)$ factor. Since we will be using this analysis only to make order-of-magnitude estimates, the exact factors are unimportant.}
\begin{equation}
H^2_\star = \pm\frac{\beta_2m^2}{\sqrt3\alpha^2} + \mathcal{O}(\alpha^0), \label{eq:b1b2hstab}
\end{equation}
where $H_\star$ is defined as the Hubble rate at the time when
$\omega^2=0$, i.e., after which the gradient instability is absent. We
pick the negative branch of \cref{eq:b1b2hstab} for physical reasons,
i.e., so that $H^2_\star>0$ given that $\beta_2<0$. We have checked
explicitly that by solving for $y$ with this value of $H$ and plugging
it into $\omega^2$, all terms up to $\mathcal{O}(\alpha^2)$ vanish. 

Interestingly, \cref{eq:b1b2hstab} is the same as the condition
\eqref{eq:backvalid} for the small-$\alpha$ expansion of the
background solution to be valid. Therefore, simply by pushing the
instability back to early times, one gets late-time bimetric dynamics
that can be described as perturbative corrections to GR, except for
the effective cosmological constant which remains nonperturbative.
This is nontrivial; while we expect everything to reduce to GR at late times when
we can expand in $\alpha H/\sqrt{\beta_n}m$, there could in principle have
been earlier times during which perturbations were stable but still
fundamentally different than in GR.

We can rewrite \cref{eq:b1b2hstab} in more physical terms as
\begin{equation}
H^2_\star = -\frac{3\sqrt3}{2\alpha^2}\left(\frac{\beta_2}{\beta_1}\right)^2H_\Lambda^2,
\end{equation}
where $H_\Lambda$ is the far-future value of $H$ and should be
comparable to the present Hubble rate, $H_0$. For $|\beta_1| \sim
|\beta_2|$, this implies simply 
\begin{equation}
H_\star \sim \frac{H_0}\alpha.
\end{equation}
We see that as we approach the GR limit, the smaller one takes the
$f$-metric Planck mass, the earlier in time bigravity's gradient
instability is cured. Our goal is to make this era so early as to be
effectively unobservable.  One has a variety of choices for the scale
where the instability sets in; the values of $\alpha$ and $\Mf$ for
various choices are summarized in \cref{tab:alpha}.

A natural requirement would be to push the instability outside the
range of the effective field theory, i.e.,~above either the cut-off
scale where new physics must enter, or the strong-coupling scale where
tree-level unitary breaks down.\footnote{These two are not always the
  same, and may not be in massive and bimetric gravity
  \cite{Aydemir:2012nz,deRham:2014zqa}.} The cut-off scale in massive
and bimetric gravity is not known. The strong-coupling scale, to
the extent it is understood, was discussed above.
\begin{table}
\begin{ruledtabular}
\begin{tabular}{l|l|l|l}
Era of transition to stability & $H_\star$ & $\alpha$ & $M_f$ \\ \hline
BBN & $10^{-16}$~eV & $10^{-17}$ & $100$~GeV \\
$\tilde\Lambda_3 = \left(m^2\Mp/\alpha\right)^{1/3}$ & $10^{-3}$~eV & $10^{-31}$ & $10^{-3}$~eV \\
GUT-scale inflation & $10^{13}$~GeV & $10^{-55}$ & $10^{-27}$~eV \\
$\Mp$ & $10^{19}$~GeV & $10^{-61}$ & $10^{-33}$~eV
\end{tabular}
\caption{The values of $\alpha$ and $\Mf$ for a few choices of the era at which perturbations become stable.}
\label{tab:alpha}
\end{ruledtabular}
\end{table}
Here we focus on observational constraints. It is natural to
demand that the instability lie beyond some important cosmic era which
we can indirectly probe, such as big-bang nucleosynthesis (BBN) or
inflation. Both of these possibilities are then likely to be
observationally safe as long as the Universe is decelerating (e.g., is
radiation-dominated) after inflation, because the instability is only
a problem for subhorizon modes with large $k/aH$, and during a
decelerating epoch modes with fixed comoving wavelength always become
smaller with respect to the horizon.  Consider, as an example, that
the transition to stability occurs between inflation and BBN. During
that period, modes will grow rapidly on small scales, but those will
be far, far smaller than the modes relevant for the cosmic microwave
background or large-scale structure.  One might worry that inflation's
ability to set initial conditions is spoiled in this scenario
(assuming that the linear theory is even valid during inflation, which
is not guaranteed due to the arguments above). However, the
instability should be absent during inflation; notice from
\cref{eq:eigenfreqs} that $\omega^2$ generically becomes large and
positive for $w$ close to $-1$.\footnote{This depends on the exact
  $\beta_n$ parameters and the evolution of $y$. Background viability
  requires $\beta_1>0$ \cite{Konnig:2013gxa}, so as long as
  $\beta_3\leq0$, at the very least the last term in
  \cref{eq:eigenfreqs} is large and manifestly positive.} Therefore
the instability would not affect the generation of primordial
perturbations during inflation. If the instability later appears with
the onset of radiation domination, it would only affect small scales
which are irrelevant for present-day cosmology.

If the instability ends at the time of BBN, $\Mf$ can be as high as
about 100~GeV, far larger than the wavenumbers probed by cosmological
observations. We remind the reader that for such a ``large'' $\Mf$, perturbations in the Einstein-Hilbert term for $f_\mn$ remain weakly-coupled for all observationally-relevant $k$.

While analytic results like \cref{eq:b1b2hstab} cannot be obtained for
most of the other two-parameter models, we have checked that in each
case the relevant behavior, $H_\star\sim H_\Lambda/\alpha$,
holds.\footnote{Specifically, this holds in the models with
  $\beta_1\neq0$. The gradient instability is absent from the
  $\beta_2\beta_3$ and $\beta_2\beta_4$ models at early times \cite{Konnig:2014xva}.
  These were shown in \rcite{Konnig:2013gxa} to have problematic
  background behavior at early times, but these again can be made
  unobservably early in the GR limit.} The values given in
\cref{tab:alpha} are therefore fairly model-independent. 

The other pathology that is typical of massive and bimetric gravity,
the Higuchi ghost, is not present
in these models. There is a simple condition for the absence of this
ghost, $\dd\rho/\dd y<0$ \cite{Yamashita:2014cra,DeFelice:2014nja} (see
also \rcite{Fasiello:2013woa,Lagos:2014lca}). Because for normal
matter $\rho$ is always decreasing with time, this amounts to
demanding that $y$ be increasing. In the ``finite-branch'' solutions
which we are considering, $y$ evolves monotonically from $0$ at early
times to a fixed positive value at late times, and so the Higuchi
bound is always satisfied \cite{Konnig:2015lfa}. 

\para{Parameter rescalings} We have presented a physically
well-motivated region of bimetric parameter space, near the GR limit,
in which observable cosmological perturbations are stable and yet
self-acceleration remains. One is naturally led to ask how this has
been missed by the many previous studies of bimetric cosmology. The
issue lies in a rescaling which leaves the action \eqref{eq:action}
invariant \cite{Berg:2012kn,Hassan:2012wr}, 
\begin{equation}
f_\mn \to \Omega^2f_\mn,\qquad \beta_n \to \frac{1}{\Omega^n} \beta_n,\qquad \Mf \to \Omega \Mf, \label{eq:rescaling}
\end{equation}
and hence gives rise to a redundant parameter. It has become common to
let $\alpha$ play this role and perform the rescaling
$\Omega=1/\alpha$ such that $\alpha$ is set to unity.  While our
results do not invalidate this rescaling, they do show that it picks
out a particular region of parameter space which may not capture all
physically-meaningful situations. In particular, the $\alpha\to0$
limit, in which the theory approaches GR---the behavior at the heart
of our removing the gradient instability---would look extremely odd
after this rescaling: the $\beta_n$ would not only be very large, but
each $\beta_{n+1}$ would be \emph{parametrically} larger than
$\beta_n$.\footnote{We can recast this as a large $m^2$, but there would remain a specific tuning among the $\beta_n$ of the form $\beta_n/\beta_{n+1}\sim\epsilon$, where $\epsilon$ is the value of $\alpha$ before the rescaling.} Therefore, studies which set $\alpha$ to unity could in
principle have found the GR-like solutions which we study here, but
only by looking at what would have appeared to be a highly unnatural
and tuned set of parameters, even though they have a simple and
sensible physical explanation. Without performing this rescaling, we
can simply take the nonzero $\beta_n$ to be $\mathcal{O}(1)$ and
consider that we are in the small-$\Mf$ r\'egime. 

It is clear that in phenomenological studies of bigravity,
$\alpha$ must not automatically be set to unity. When working with a
two-$\beta_n$ model, perhaps a more sensible rescaling would be one
such that the two $\beta_n$ are equal to each other (up to a possible
sign). They can further be absorbed into $m^2$. In this case, the free parameters are effectively the spin-2
interaction scale, $m^2$, and the $f$-metric Planck mass,
$\Mf$. Their effects decouple nicely: $\Mf$ controls the earliness of
the instability, while $m$ sets the acceleration scale.
Alternatively, one may consider that the rescaling
\eqref{eq:rescaling} simply tells us that rather different regions of
parameter space happen to have the same solutions, and therefore not
perform any rescaling \textit{a priori} at all.

\para{Summary and discussion} We have shown that a well-motivated but
heretofore underexplored region of parameter space in bimetric gravity
can lead to cosmological solutions which are observationally viable
and close to general relativity, with an effective cosmological
constant that is set by the spin-2 interaction scale $m$. In this
limit, obtained by taking a small $f$-metric Planck mass, the gradient
instability that seems to generically plague bimetric models at late
times is relegated to the very early Universe, where it can be either
made unobservable or pushed outside the r\'egime of validity of the
effective theory. This instability had been considered in previous
work to make bimetric cosmologies nonpredictive even at late
times. Furthermore, in this limit the theory avoids the usual low-scale strong-coupling issue that affects the helicity-0 sector in the massive-gravity limit.

What is encouraging is that the one property of bigravity which
survives in the small-$\alpha$ limit is its cosmologically most useful
feature, the technically-natural dark energy scale. In other words,
the effective cosmological constant of bigravity in a region close to
GR is not just the vacuum-energy contribution and can give rise to
self-acceleration in its absence.

The model we have presented is expected to be extremely close to GR at
all but very high energy scales. In particular the Newtonian limit is
well-behaved; unlike $m^2\to0$, which suffers from the vDVZ discontinuity, the GR limit $\alpha\to0$ is
completely smooth because all the helicity states of the massive
spin-2 mode decouple from matter. Note also that massive gravity does
not possess such a continuous GR limit. 

It is worth emphasizing that the $\alpha\rightarrow 0$ limit brings
bimetric theory arbitrarily close to GR even for a large value of the
spin-2 mass scale, $m\gg H_0$. The presence of heavy spin-2
fields in the Universe is therefore not excluded as long as their
self-interaction scale (set by $M_f$) is sufficiently small compared
to $M_\mathrm{Pl}$. In this case, however, the $\beta_n$ parameters need to be highly tuned for the effective cosmological constant small enough to be compatible with observations.\footnote{Indeed, without this tuning of the $\beta_n$, the interaction term would lead to acceleration at an unacceptably early epoch. This scenario is related to the findings of \rcite{DeFelice:2014nja}, where it was shown that the instability becomes negligible for large values of $m$.} Note however that, since the $\beta_n$ are protected against loop corrections \cite{Hinterbichler:2010xn,deRham:2012ew,deRham:2013qqa}, this tuning does not violate \emph{technical} naturalness.

Finally we comment on the potential observable signatures of this theory. While at low energies, corresponding to recent cosmological epochs, this limit of bigravity is extremely close to GR, there may be observable effects at early times when the effects of strong coupling become important. In this case, given by $H>H_\star$, the small-$\alpha$ approximation breaks down and modified-gravity effects must be taken into account. This may be particularly important for inflation, which will see such effects unless $\Mf$ is extraordinarily small. A better understanding of strong coupling in the $f_\mn$ sector will therefore point the way towards tests of this important region of bimetric parameter space, since at this point it is not clear how to perform computations in the strong-coupling regime. There may also be effects related to the Vainshtein mechanism \cite{Vainshtein:1972sx,Babichev:2013usa}. We conclude that the closeness of this theory to GR is both a blessing and a curse: while it is behind the exorcism of the gradient instability and brings the theory in excellent agreement with experiments, it presents a serious observational challenge if it is to be compared \emph{against} GR. It is nevertheless encouraging that this ``GR-adjacent'' bigravity naturally explains cosmic acceleration while avoiding the instabilities that plague other bimetric models, and therefore merits serious consideration.

\begin{acknowledgments}
We are grateful to Luca Amendola, Jonas Enander, Matteo Fasiello, Emir G\"umr\"uk\c c\"uo\u glu, Nima Khosravi, Edvard M\"ortsell, Luigi Pilo, Marit Sandstad, and Sergey Sibiryakov for useful discussions. We thank the ``Extended Theories of Gravity'' workshop at Nordita for providing a stimulating atmosphere during the completion of this work. Y.A. and F.K. acknowledge support from DFG through the project TRR33 ``The Dark Universe.'' F.K. is also supported by the Graduate College ``Astrophysics of Fundamental Probes of Gravity.'' The work of A.S.M. is supported by ERC grant No. 615203 under the FP7 and the Swiss National Science Foundation through the NCCR SwissMAP. A.R.S. acknowledges support from the STFC.
\end{acknowledgments}

\bibliography{bibliography}

\end{document}